\documentclass{llncs}
\usepackage[table]{xcolor}
\usepackage{colortbl}
\usepackage{llncsdoc}
\usepackage{graphicx}
\usepackage{amsmath,lipsum}
\usepackage{amssymb}
\usepackage{float}
\usepackage{hhline}
\usepackage{tikz,tkz-base}
\usetikzlibrary{shapes,decorations,shadows}
\usetikzlibrary{decorations.pathmorphing}
\usetikzlibrary{decorations.shapes}
\usetikzlibrary{fadings}
\usetikzlibrary{patterns}
\usetikzlibrary{calc}
\usetikzlibrary{decorations.text}
\usetikzlibrary{decorations.footprints}
\usetikzlibrary{decorations.fractals}
\usetikzlibrary{shapes.gates.logic.IEC}
\usetikzlibrary{shapes.gates.logic.US}
\usetikzlibrary{fit,chains}
\usetikzlibrary{positioning}
\usepgflibrary{shapes}
\usetikzlibrary{scopes}
\usetikzlibrary{arrows}
\usetikzlibrary{backgrounds}
\usetikzlibrary{decorations.pathreplacing,calc}

\tikzset{latent/.style={circle,fill=white,draw=black,inner sep=1pt, 
minimum size=20pt, font=\fontsize{10}{10}\selectfont},
obs/.style={latent,fill=gray!25},
const/.style={rectangle, inner sep=0pt},
factor/.style={rectangle, fill=black,minimum size=5pt, inner sep=0pt},
>={triangle 45}}
\begin{document}
\title{Protein (Multi-)Location Prediction: Using Location
Inter-Dependencies in a Probabilistic Framework}

\author{Ramanuja Simha\inst{1} \and Hagit Shatkay\inst{1,2,3}}

\institute{Department of Computer and Information Sciences, \\ University of Delaware, Newark, DE, USA
\and
Center for Bioinformatics and Computational Biology, DBI, \\ University of Delaware, Newark, DE, USA
\and
School of Computing, Queen's University, Kingston, ON, Canada}

\maketitle
\begin{abstract}
Knowing the location of a protein within the cell is important for understanding its function, role in biological processes, and potential use as a drug target. Much progress has been made in developing computational methods that predict single locations for proteins, assuming that proteins localize to a single location. However, it has been shown that proteins localize to multiple locations. While a few recent systems have attempted to predict multiple locations of proteins, they typically treat locations as independent or capture inter-dependencies by treating each locations-combination present in the training set as an individual location-class. We present a new method and a preliminary system we have developed that directly incorporates inter-dependencies among locations into the multiple-location-prediction process, using a collection of Bayesian network classifiers. We evaluate our system on a dataset of single- and multi-localized proteins. Our results, obtained by incorporating inter-dependencies are significantly higher than those obtained by classifiers that do not use inter-dependencies. The performance of our system on multi-localized proteins is comparable to a top performing system (YLoc$^{+}$), without restricting predictions to be based only on location-combinations present in the training set.
\end{abstract}
\section{Introduction}
\label{sec:intro}
Knowing the location of a protein within the cell is essential for understanding its function, its role in biological processes, as well as its potential role as a drug target~\cite{alberts-et-al:b:mol-bio,rost-et-al:j:auto-pred-prot,bakheet-et-al:j:prot-drug-tar}. Experimental methods for protein localization such as those based on mass spectrometry~\cite{dreger:j:proteome-struc} or green fluorescence detection~\cite{simpson:j:sys-cdna,hanson-et-al:j:gfp}, although often used in practice, are time consuming and typically not cost-effective for high-throughput localization. Hence, an ongoing effort is put into developing high-throughput computational methods~\cite{nakai-et-al:j:prot-loc-gramneg,emanuelsson-et-al:j:prot-loc-aa,rey-et-al:j:prec-bac-high,shatkay-et-al:j:sherloc,blum-et-al:j:multiloc2-go} to obtain proteome-wide location predictions.

Over the last decade, there has been significant progress in the development of computational methods that predict a {\em single} location per protein. The focus on single-location prediction is driven both by the data available in public databases such as UniProt, where proteins are typically assigned a single location, as well as by an (over-)simplifying assumption that proteins indeed localize to a single location. However, proteins do localize to multiple compartments in the cell~\cite{foster-et-al:j:mam-org-map,zhang-et-al:j:dbmloc,millar-et-al:j:plant-prot,murphy:j:comm-sub-dis}, and translocate from one location to another~\cite{pohlschroder-et-al:j:div-prot-trans}. Identifying the mutiple locations of a protein is important because translocation can serve some unique functions. For instance, GLUT4, an insulin-regulated glucose transporter, which is stored in the intracellular vesicles of adipocytes, translocates to the plasma membrane in response to insulin~\cite{rea-et-al:j:glut4,russell-et-al:j:glut4}. As proteins do not localize at random and translocations happen between designated inter-dependent locations, we hypothesize that modeling such inter-dependencies can help in predicting protein locations. Thus, we aim to identify associations or \emph{inter-dependencies} among locations and leverage them in the process of predicting locations for proteins.

Several methods have been recently suggested for predicting multiple locations for proteins. ngLOC~\cite{king-et-al:j:ngloc} uses a Na\"{i}ve Bayes classifier to obtain {\it independent predictions} for each single location and combines these individual predictions to obtain a multi-location prediction. Li et al.~\cite{li-et-al:j:prot-multi-pseudo} construct multiple binary classifiers, each using an ensemble of $k$-nearest neighbor and SVM, where each binary classifier distinguishes between a pair of locations. The predictions from all the classifiers are combined to obtain a multi-location prediction. iLoc-Euk~\cite{chou-et-al:j:iloc-euk} uses a multi-label $k$-nearest neighbor classifier to predict multiple locations for proteins. Similar methods were used for localizing subsets of eukaryotic proteins~\cite{chou-et-al:j:iloc-hum,wu-et-al:j:iloc-plant}, virus proteins~\cite{xiao-et-al:j:iloc-virus}, and bacterial proteins~\cite{xiao-et-al:j:gram-neg,wu-et-al:j:iloc-gpos}. In contrast to the machine learning-based approaches listed above, KnowPred~\cite{lin-et-al:j:knowpred} uses sequence similarity to associate proteins with multiple locations. 

Notably, none of the above methods for predicting multiple locations utilizes inter-dependencies among locations in the prediction process. All the above models independently predict each single location and thus do not take into account predictions for other locations. IMMML~\cite{he-et-al:j:immml} attempts to make use of {\it correlation} among pairs of locations, a simple type of dependency, when predicting multiple locations for proteins. This system does not account for more complex inter-dependencies and was not tested on any extensive protein multi-localization dataset. YLoc$^{+}$\cite{briesemeister-et-al:j:yloc}, a comprehensive system for protein location prediction, uses a na\"{i}ve Bayes classifier (see e.g. \cite{mitchell:b:ml}) and captures protein localization to multiple locations by explicitly \emph{introducing a new class for each combination of locations supported by the training set} (i.e. having proteins localized to the combination). Thus, each prediction performed by the na\"{i}ve Bayes classifier can assign a protein to only those combinations of locations included in the training data. To produce its output, YLoc$^{+}$ transforms the prediction into a multinomial distribution over the individual locations. We also note that as the number of possible location-combinations is exponential in the number of locations, training the na\"{i}ve Bayes classifier in this manner does not provide a practical model in the general case of multi-localized proteins, beyond the training set. The performance of YLoc$^{+}$ was evaluated using an extensive dataset~\cite{briesemeister-et-al:j:yloc} and is the highest among current multi-location predictors.

In this paper, we present a new method that directly models inter-dependencies among locations and incorporates them into the process of predicting locations for proteins. Our system is based on a collection of Bayesian network classifiers~\cite{grossman-et-al:c:bnc}. Each Bayesian Network (BN) related to each classifier corresponds to a single location $L$. Each such network is used to assign a probability for a protein to be found at location $L$, given both the protein's features and {\it information regarding the protein's other possible locations}. Learning each BN involves learning the dependencies among the other locations that are primarily related to proteins localizing to location $L$. For each Bayesian network classifier, its corresponding BN is learnt with the goal to improve the classifier's prediction quality. The formulation of multi-location prediction as classification via Bayesian networks, as well as the network model are presented in Section~\ref{sec:prob-form}. Notably, our system does not assume that {\it all} proteins it classifies  are multi-localized, but rather more realistically, that proteins may be assigned to one or more locations. 

We train and test our preliminary system on a dataset containing single- and multi-localized proteins previously used in the development and testing of the YLoc$^{+}$ system~\cite{briesemeister-et-al:j:yloc}, which includes the most comprehensive collection of multi-localized proteins currently available, derived from the DBMLoc dataset~\cite{zhang-et-al:j:dbmloc}. As done in other studies~\cite{shatkay-et-al:j:sherloc,blum-et-al:j:multiloc2-go,briesemeister-et-al:j:yloc,hoeg-et-al:j:multiloc}, we use multiple runs of 5-fold cross-validation. The results clearly demonstrate the advantage of using location inter-dependencies. The $F_1$ score of 81\% and overall accuracy of 76\% obtained by incorporating inter-dependencies are significantly higher than the corresponding values obtained by classifiers that do not use inter-dependencies. Also, while our system retains a level of performance comparable to that of YLoc$^{+}$ on the same dataset, we note that unlike YLoc$^{+}$, by training the individual classifiers to predict individual $-$ although inter-dependent $-$ locations, the training of our system is not restricted to only those combinations of locations present in the dataset, thus our system is generalizable to multi-locations beyond those included in the training set.  

The rest of the paper proceeds as follows: Section~\ref{sec:prob-form} formulates the problem of protein subcellular multi-location prediction and briefly provides background on Bayesian networks and relevant notations. Section~\ref{sec:methods} discusses the structure, parameters, and inter-dependencies comprising our Bayesian network collection, and introduces the learning procedure used for finding them. Section~\ref{sec:results} presents details of the dataset, the performance evaluation measures, and experimental results. Section~\ref{sec:discussion} summarizes our findings and outlines future directions.
\section{Problem Formulation}
\label{sec:prob-form}
As is commonly done in the context of classification, and protein-location classification in particular~\cite{emanuelsson-et-al:j:prot-loc-aa,blum-et-al:j:multiloc2-go,briesemeister-et-al:j:yloc,garg-et-al:j:eslpred2}, we represent each protein, $P$, as a weighted feature vector, \mbox{$\vec{f^P} \!\!=\!\! \langle f_{1}^P, \ldots, f_{d}^P \rangle$}, where $d$ is the number of features. We view each feature as a random variable $F_i$ representing a characteristic of a protein, such as the presence or absence of a short amino acid motif~\cite{emanuelsson-et-al:j:prot-loc-aa,hoeg-et-al:j:multiloc}, the relative abundance of a certain amino acid as part of amino-acid composition~\cite{king-et-al:j:ngloc}, or the annotation by a Gene Ontology (GO) term~\cite{huang-et-al:j:proloc-go}. Each vector-entry, $f_i^P$, corresponds to the value taken by feature $F_i$ with respect to protein $P$. In the experiments described here, we use the exact same representation used by Briesemeister et al.~\cite{briesemeister-et-al:j:yloc} as explained in Section~\ref{subsec:data-prep}. 

We next introduce notations relevant to the representation of a protein's localization. Let $S \!=\! \{s_1, \ldots, s_q\}$ be the set of $q$ possible subcellular components in the cell. For each protein $P$, we represent its location(s) as a vector of 0/1 values indicating the protein's absence/presence, respectively, in each subcellular component. The location-indicator vector for protein $P$ is thus a vector of the form: $\vec{l^P} \!=\! \langle {l_{1}^P}, \ldots, {l_{q}^P} \rangle$ where ${l_{i}^P} \!=\! 1$ if $P$ localizes to $s_i$ and $l_{i}^P \!=\! 0$ otherwise. As with the feature values, each location value, $l_i^P$ is viewed as the value taken by a random variable, where for each location, $s_i$, the corresponding random variable is denoted by  $L_i$. Given a dataset consisting of $m$ proteins along with their location vectors, we denote the dataset as: $D \!=\! \{(P_j, \vec{l}^{\;P_j})~|~1 \leq j \leq m\}$. We thus view the task of protein subcellular multi-location prediction as that of developing a classifier (typically learned from a dataset $D$ of proteins whose locations are known) that given a protein $P$ outputs a $q$-dimensional location-indicator vector that represents $P$'s localization.

As described in Section~\ref{sec:intro}, most recent approaches that extend location-prediction beyond a single location (e.g. KnowPred~\cite{lin-et-al:j:knowpred}, and Euk-mPLoc 2.0~\cite{chou-et-al:j:eukmploc20}), do not consider inter-dependencies among locations. YLoc$^{+}$\cite{briesemeister-et-al:j:yloc} indirectly considers these inter-dependencies by creating a class for each location-combination. Our underlying hypothesis, which is supported by the experiments and the results presented here, is that capturing location inter-dependencies directly can form the basis for a generalizable approach for location-prediction. The training of a classifier for protein multi-location prediction involves learning these inter-dependencies so that the classifier can leverage them in the prediction process. We use  Bayesian networks to model such inter-dependencies.

In order to develop a protein subcellular multi-location predictor, we propose to develop a collection of classifiers, $C_1, \ldots, C_q$, where the classifier $C_i$ is viewed as an ``expert'' responsible for predicting the 0/1 value, $l_{i}^P$, indicating $P$'s non-localization or localization to $s_i$. In order to make use of location inter-dependencies, each $C_i$ uses estimates of location indicators of $P$, $\hat{l}_{j}^P$ (for all other locations $j$, where $j \neq i$), along with the feature-values of $P$, in order to calculate a prediction. We use support vector machines (SVMs) (see e.g.~\cite{mitchell:b:ml}) to compute these estimates. The output of $C_i$ for a protein $P$ is given by 
\begin{align}
\hspace{-0.5cm}C_i(P) = 
\begin{cases} 
1 & \mbox{If } \Pr(l_{i}^P=1~|~P,\hat{l}_{1}^P, \ldots, \hat{l}_{i-1}^P, \hat{l}_{i+1}^P, \ldots, \hat{l}_{q}^P) > 0.5 \mbox{;} \\ 
0 & \mbox{Otherwise.}
\end{cases}  \label{eqn:thresholding}
\end{align}
Further details about the estimation procedure itself are provided in Section~\ref{subsec:infer-pred}. 

Bayesian networks have been used before in many biological applications (e.g. \cite{friedman-et-al:j:bn-exprdata,segal-et-al:j:prob-geneexp,lee-et-al:j:bntagger}). In this paper, we use them to model inter-dependencies among subcellular locations, as well as among protein-features and locations. We briefly introduce Bayesian networks here, along with the relevant notations (see~\cite{jensen-et-al:b:bn} for more details). A Bayesian network consists of a directed acyclic graph $G$, whose nodes are random variables, which in our case represent features, denoted $F_1, \ldots, F_d$, and location indicators, denoted $L_1, \ldots, L_q$. We assume here that all the feature values are discrete. To ensure that, we use the recursive minimal entropy partitioning technique~\cite{fayyad-et-al:c:disc-class} to discretize the features; this technique was also used in the development of YLoc$^{+}$\cite{briesemeister-et-al:j:yloc}.

\begin{figure}[t]
\vspace*{-0.55cm}
\noindent \begin{centering}
\begin{tikzpicture}
\node [matrix,matrix anchor=mid, column sep=8pt, row sep=2pt] {
& \node () [] {\pmb{$C_1$}}; & & & & & & & & \node () [] {\pmb{$C_q$}}; & \\ 
\node (m1f1) [latent,fill=gray!50,scale=0.7] {$F_1$}; & & \node (m1l2) [latent,fill=gray!50,scale=0.7] {$L_2$}; & & &  \node (mklk) [] {$\ldots$}; & & & \node (m2f1) [latent,fill=gray!50,scale=0.7] {$F_1$}; & &  \node (m2l1) [latent,fill=gray!50,scale=0.7] {$L_1$}; \\
\node (m1f2) [latent,fill=gray!50,scale=0.7] {$F_2$}; & & \node (m1l3) [latent,fill=gray!50,scale=0.7] {$L_3$}; & & & & & & \node (m2f2) [latent,fill=gray!50,scale=0.7] {$F_2$}; & & \node (m2l2) [latent,fill=gray!50,scale=0.7] {$L_2$}; \\
\node [] {$\vdots$}; & & \node [] {$\vdots$}; & & & && & \node [] {$\vdots$}; & & \node [] {$\vdots$}; \\
\node (m1fd) [latent,fill=gray!50,scale=0.7] {$F_d$}; & \node [] {$\vdots$}; & \node (m1lq) [latent,fill=gray!50,scale=0.7] {$L_q$}; & & & & & & \node (m2fd) [latent,fill=gray!50,scale=0.7] {$F_d$}; & \node [] {$\vdots$}; & \node (m2lq-1) [latent,fill=gray!50,scale=0.7] {$L_{q\!-\!\!1}$};\\
& \node (m1l1) [latent,scale=0.7] {$L_1$}; & & & & \node (mklk) [] {$\ldots$}; & & & & \node (m2lq) [latent,scale=0.7] {$L_q$}; & \\ 
};
\draw [->] (m1f1) -- (m1l2) ;
\draw [<-] (m1f2) -- (m1lq) ;
\draw [->] (m1f1) -- (m1l3) ;
\draw [->] (m1f2) -- (m1l1) ;
\draw [->] (m1l1) -- (m1l3) ;
\draw [->] (m1f1) -- (m1lq) ;
\draw [->] (m1fd) -- (m1l1) ;
\draw [->,solid,black] (m1lq) to[out=0,in=0] (m1l3) ;
\draw [->] (m2f1) -- (m2l1) ;
\draw [<-] (m2f2) -- (m2lq-1) ;
\draw [<-] (m2fd) -- (m2l2) ;
\draw [->] (m2f1) -- (m2l2) ;
\draw [->] (m2f1) -- (m2lq-1) ;
\draw [->] (m2fd) -- (m2lq) ;
\draw [->] (m2lq-1) -- (m2lq) ;
\draw [->,solid,black] (m2l2) to[out=0,in=0] (m2lq-1) ;
\end{tikzpicture}
\par\end{centering}
\vspace*{-0.4cm}
\caption{{\footnotesize An example of a collection of Bayesian network classifiers we learn. 
The collection consists of several classifiers $C_1, \ldots, C_q$, one for each of the $q$ subcellular locations. Directed edges represent dependencies between the connected nodes. We note that there are edges among location variables $(L_1, \ldots, L_q)$, as well as between feature variables ($F_1, \ldots, F_d$) and location variables ($L_1, \ldots, L_q$), but not among the feature variables. The latter indicates independencies among features, as well as conditional independencies among features given the locations.}
\label{fig:ebnc}}
\vspace*{-0.65cm}
\end{figure}
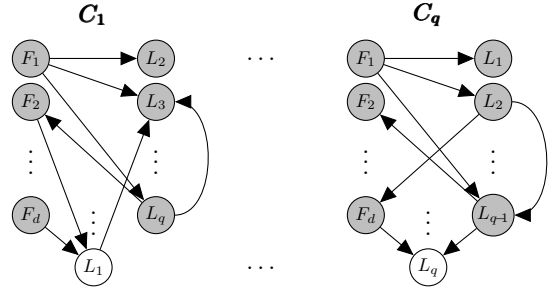

Directed edges in the graph indicate inter-dependencies among the random variables. Thus, as demonstrated in Figure~\ref{fig:ebnc}, edges are allowed to appear between feature- and location-nodes, as well as between pairs of location-nodes in the graph. Edges between location-nodes directly capture the inter-dependencies among locations. We note that there are no edges between feature-nodes in our model, which reflects an assumption that features are either independent of each other or conditionally independent given the locations. This simplifying assumption helps speed up the process of learning the network structure from the data, while the other allowed inter-dependencies still enable much of the structure of the problem to be captured (as demonstrated in the results). Further details about the learning procedure itself are provided in Section~\ref{subsec:learn-bncs}. 

To complete the Bayesian network framework, each node \linebreak $v \!\in\! \{\!F_1\!, \ldots\!, F_d, L_1\!, \ldots\!, L_q\!\}$ in the graph is associated with a conditional probability table, $\theta_v$, containing the conditional probabilities of the values the node takes given its parents' values, $\Pr(v~|~Pa(v))$. We denote by $\Theta$ the set of all conditional probability tables, and the Bayesian network is the pair \mbox{$(G, \Theta)$}. 
A consequence of using the Bayesian network structure, is that it represents certain conditional independencies among non-neighboring nodes~\cite{jensen-et-al:b:bn}, such that the joint distribution of the set of network variables can be simply calculated as:
\begin{align}
\textstyle{\Pr(F_1, \ldots, F_d, L_1, \ldots, L_q) = \prod_{i=1}^{d} \Pr(F_i~|~Pa(F_i)) \prod_{j=1}^{q} \Pr(L_j~|~Pa(L_j)) \label{eqn:bn}}.
\end{align}

Figure~\ref{fig:ebnc} shows an example of a collection of Bayesian network classifiers. The collection consists of Bayesian network classifiers $C_1, \ldots, C_q$, one for each of the $q$ subcellular locations $s_1, \ldots, s_q$, where each classifier $C_i$ consists of the graph $G_i$ and its set of parameters $\Theta_i$. In each classifier $C_i$, the location indicator variable $L_i$ is the variable we need to predict and is therefore viewed as \emph{unobserved}, and is shown as an unshaded node in the figure. The feature variables $F_1, \ldots, F_d$ are given for each protein and as such are viewed as known or \emph{observed}, shown as shaded nodes in the figure. Finally, the values for the location indicator variables for all locations except for $L_i$,  $\{L_1, \ldots, L_q\} - \{L_i\}$, are needed for calculating the predicted value for $L_i$ in the classifer $C_i$. As such, they are viewed by the classifier as though they are \emph{observed}. Notably, the values of these variables are not known and still need to be estimated. 

Thus, the structure and parameters of the network for each classifier $C_i$ (learnt as described in Section~\ref{subsec:learn-bncs}), are used to predict the value of each unobserved variable, $L_i$. The task of each classifier $C_i$, is to predict the value of the variable $L_i$ given the values of all other variables $F_1, \ldots, F_d$, and \mbox{$\{\!L_1\!, \ldots\!, L_q\!\} \!-\! \{\!L_i\!\}$}. Since, as noted above, the values of the location indicator variables $L_j$ ($j \neq i$) are unknown at the point when $L_i$ needs to be calculated, we \emph{estimate} their values, using simple SVM classifiers as described in Section~\ref{subsec:learn-bncs}. We note that other methods, such as expectation maximization, can be used to estimate all the hidden parameters, which we shall do in the future.  
\section{Methods}
\label{sec:methods}
As our goal is to assign locations (possibly multiple) to proteins, we use a collection of Bayesian network classifiers, where each classifier $C_i$, predicts the value (0 or 1) of a single location variable $L_i$  -- while using estimates of all the other location variables $L_j$ ($j \neq i$), which are assumed to be known, as far as the classifier $C_i$ is concerned. The estimates of the location values $L_j$ are calculated using SVM classifiers as described in Section~\ref{subsec:learn-bncs}. The individual predictions from all the classifiers are then combined to produce a multi-location prediction. For each location $s_i$, a Bayesian network classifier $C_i$ must be learned from training data before it can be used. As described in Section~\ref{sec:prob-form}, each classifier $C_i$ consists of a graph structure $G_i$ and a set of conditional probability parameters, $\Theta_i$, that is: \mbox{$C_i \!= \!(G_i,\Theta_i)$}. Thus, our first task is to learn the individual classifiers, i.e. their respective Bayesian network structures and parameters. The individual networks can then be used to predict a protein's localization to each location. 

Given a protein $P$, each classifier $C_i$ needs to accurately predict the location indicator value $l_{i}^P$, given the feature-values of $P$ and estimates of all the other location indicator values $\hat{l}_{j}^P$ (where $j \neq i$). That is, each classifier $C_i$ in the collection assumes that the estimates of the location-indicator values, $\hat{l}^P_j$ for all other locations $s_j$ (where $j \neq i$) are already known, and is responsible for predicting only the indicator value ${l}^P_i$ for location $s_i$, given all the other indicator values. For a Bayesian network classifier this means calculating the conditional probability 
\begin{align}
\Pr(l_{i}^P=1~|~P,\hat{l}_{1}^P, \ldots, \hat{l}_{i-1}^P, \hat{l}_{i+1}^P, \ldots, \hat{l}_{q}^P), \label{eqn2}
\end{align}
under classifier $C_i$, where $\hat{l}_{1}^P, \ldots, \hat{l}_{i-1}^P, \hat{l}_{i+1}^P, \ldots, \hat{l}_{q}^P$ are all estimated using simple SVM classifiers. The classifiers $C_1, \ldots, C_q$ are each learned by directly optimizing an objective function that is based on such conditional probabilities, calculated with respect to the training data as explained in Section~\ref{subsec:learn-bncs}.

The procedures used for learning the Bayesian network classifiers and to combine the individual network predictions are described throughout the rest of the section.  
\subsection{Structure and Parameter Learning of Bayesian Network Classifiers}
\label{subsec:learn-bncs}
Given a dataset $D$, consisting of a set of $m$ proteins $\{P_1, \ldots, P_m\}$ and their respective location vectors $\{\vec{l}^{P_1}, \ldots, \vec{l}^{P_m}\}$, each classifier $C_i$ is trained so as to produce the ``best'' prediction possible for the value of the location indicator ${l}^{P}_i$ (for location $s_i$), for any given protein $P$ and a set of estimates of location indicators for all other locations (as shown in Equation~\ref{eqn2} above). Based on this aim and on the available training data, we use the \emph{Conditional Log Likelihood (CLL)} as the objective function to be optimized when learning each classifier $C_i$. Classifiers whose structures were learnt by optimizing this objective function were found to perform better than classifiers that used other structures~\cite{grossman-et-al:c:bnc}. This objective function is defined as: 
\begin{align*}
CLL(C_{i}~|~D) = \sum_{j=1}^{m} \log \Pr(L_i = {l}^{P_j}_i~|~\vec{f}^{P_j}, {\hat{l}^{P_j}}_1, \ldots, {\hat{l}^{P_j}}_{i-1}, {\hat{l}^{P_j}}_{i+1}, \ldots, {\hat{l}^{P_j}}_q). 
\end{align*}
Each $P_j$ is a protein in the training set and each probability term is the conditional probability of protein $P_j$ to have the indicator value ${l}^{P_j}_i$ (for location $s_i$), given its feature vector $\vec{f}^{P_j}$ and the current estimates for all the other location indicators are $\hat{l}^{P_j}_k$ (where $k \neq i$), under the Bayesian network structure $G_i$ for the classifier $C_i$ that governs the joint distribution of all the variables in the network (see Equation~\ref{eqn:bn}). 

To learn a Bayesian network classifier that optimizes this objective function, we use a greedy hill climbing search (see~\cite{grossman-et-al:c:bnc,heckerman-et-al:b:learn-bn} for details). While Grossman and Domingos~\cite{grossman-et-al:c:bnc} propose a heuristic method that modifies the basic search depicted by Heckerman et al.~\cite{heckerman-et-al:b:learn-bn}, we do not employ it in this preliminary study, but rather use the basic search, as it does not prove to be prohibitively time consuming. To find estimates for the location indicator values $\hat{l}^{P_j}_k$, we compute a one-time estimate for each indicator ${l}^{P_j}_i$ from the feature-values of the protein $\vec{f}^{P_j}$ by using an SVM classifier (e.g.~\cite{mitchell:b:ml}). We use the SVM implementation provided by the Scikit-learn library~\cite{pedregosa-et-al:j:scikit} with a Radial Basis Function kernel. We employ $q$ such SVMs, $\text{\it{SVM}}_1, \ldots, \text{\it{SVM}}_q$, where each SVM classifier is trained to distinguish one location indicator from the rest, as done in the Binary Relevance approach~\cite{tsoumakas:j:mulilabel-overview}. The rest of the network parameters are estimated as follows: For each Bayesian network classifier $C_i$, we use the maximum likelihood estimates calculated from frequency counts in the training dataset, $D$, to estimate the network parameters (see~\cite{grossman-et-al:c:bnc}). To avoid overfitting of the parameters, we apply standard smoothing by adding pseudo-counts for all the events that have zero counts (see~\cite{russell-et-al:b:ai} for details).

To summarize, at the end of the learning process we have $q$ Bayesian network classifiers, $C_1, \ldots, C_q$, like the ones depicted in Figure~\ref{fig:ebnc}, and $q$ SVMs, $\text{\it{SVM}}_1, \ldots, \text{\it{SVM}}_q$, used for obtaining initial estimates for each location variable for any given protein. We next describe how these classifiers are used to predict the multi-location of a protein $P$.
\subsection{Multiple Location Prediction}
\label{subsec:infer-pred}
Given a protein $P$, whose locations we would like to predict, we first use the SVMs to obtain preliminary estimates for each of its location indicator values $\hat{l}^{P}_1, \ldots, \hat{l}^{P}_q$. We then use each of the learned classifiers $C_i$, and the preliminary values obtained from the SVMs to predict the value of the location indicator $l_{i}^P$. The classifier outputs a value of either a 0 or a 1 by thresholding, as shown in Equation~\ref{eqn:thresholding}. The conditional probability of $l_{i}^P$ given the feature-values of the protein $P$ and the estimates of the location indicator values $\hat{l}_{j}^P$ (where $j \neq i$) is first calculated as:
\begin{align}
\Pr(l_{i}^P=1~|&~\vec{f^P},\hat{l}_{1}^P, \ldots, \hat{l}_{i-1}^P, \hat{l}_{i+1}^P, \ldots, \hat{l}_{q}^P) = \nonumber \\ 
& \frac{\Pr(l_{i}^P=1,\vec{f^P}, \hat{l}_{1}^P, \ldots, \hat{l}_{i-1}^P, \hat{l}_{i+1}^P, \ldots, \hat{l}_{q}^P)}{\sum_{z \in \{0,1\}}\Pr(l_{i}^P=z,\vec{f^P}, \hat{l}_{1}^P, \ldots, \hat{l}_{i-1}^P, \hat{l}_{i+1}^P, \ldots, \hat{l}_{q}^P)}\label{eqn:inf}. 
\end{align}
The joint probabilities in the numerator and the denominator of Equation~\ref{eqn:inf} above are factorized into conditional probabilities using the Bayesian network structure, $G_i$ (see Equation~\ref{eqn:bn}). The 0/1 prediction for each $l_{i}^P$ obtained from each $C_i$ becomes the value of the $i$'th position in the location-indicator vector $\langle l_1^P, \ldots, l_q^P \rangle$ for protein $P$. This is the total multi-location prediction for protein $P$. 

In the next section, we describe our experiments using the Bayesian network framework for predicting protein multi-location and the results obtained. 
\section{Experiments and Results}
\label{sec:results}
We implemented our algorithms for learning and using a collection of Bayesian network classifiers as described above using Python and the machine learning library Scikit-learn~\cite{pedregosa-et-al:j:scikit}. We have applied it to a dataset containing single- and multi-localized proteins, previously used for training YLoc$^{+}$~\cite{briesemeister-et-al:j:yloc}. Below we describe the dataset, the experiments, the evaluation methods we use, and the multiple location prediction results obtained on the proteins from this dataset. 
\subsection{Data Preparation}
\label{subsec:data-prep}
In our experiments we use a dataset containing 5447 single localized proteins (originally published as the H{\"o}glund dataset~\cite{hoeg-et-al:j:multiloc}) and 3056 multi-localized proteins (originally published as part of the DBMLoc set~\cite{zhang-et-al:j:dbmloc} that is no longer publicly available). The combined dataset was previously used by Briesemeister et al.~\cite{briesemeister-et-al:j:yloc} in their extensive comparison of multi-localization prediction systems. 
We report results obtained over the multi-localized proteins for comparing our system to other published systems, since the results for these systems are only available for this subset~\cite{briesemeister-et-al:j:yloc}. For all other experiments described here, we report results obtained over the combined set of single- and multi-localized proteins.
We use the exact same representation of a 30-dimensional feature vector as used in YLoc$^{+}$~\cite{briesemeister-et-al:j:yloc}. The features include sequence-based features, e.g. amino acid composition and those based on PROSITE patterns, as well as on GO annotations. (See~\cite{briesemeister-et-al:j:yloc} for details on the pre-processing, feature construction, and feature selection). The single localized proteins are from the following locations (abbreviations and number of proteins per location is given in parentheses):
cytoplasm (\emph{cyt}, 1411 proteins); endoplasmic reticulum (\emph{ER}, 198), extra cellular space (\emph{ex}, 843), golgi apparatus (\emph{gol}, 150), lysosomal (\emph{lys}, 103), mitochondrion (\emph{mi}, 510), nucleus (\emph{nuc}, 837), membrane (\emph{mem}, 1238), and peroxisomal (\emph{per}, 157). 
The multi-localized proteins are from the following pairs of locations: cyt\_nuc (1882 proteins), ex\_mem (334), cyt\_mem (252), cyt\_mi (240), nuc\_mi (120), ER\_ex (115), and ex\_nuc (113). Note that all the multi-location subsets used have over 100 representative proteins. 
\subsection{Experimental Setting and Performance Measures}
To compare the performance of our system to that of other systems (YLoc$^+$~\cite{briesemeister-et-al:j:yloc}, Euk-mPLoc~\cite{chou-et-al:j:eukmploc}, WoLF PSORT~\cite{horton-et-al:j:wolfpsort}, and KnowPred~\cite{lin-et-al:j:knowpred}), whose performance on a large set of multi-localized proteins was described in a previously published comprehensive study~\cite{briesemeister-et-al:j:yloc}, we use the exact same dataset, employing the commonly used stratified 5-fold cross-validation. As the information about the exact 5-way splits used before is not available, we ran five complete runs of 5-fold-cross-validation (i.e. 25 runs in total), where each complete run of 5-fold cross-validation uses a different 5-way split. The use of multiple runs with different splits helps validate the stability and the significance of the results. To ensure that the results obtained by using our 5-way splits for cross-validation can be fairly compared with those reported before~\cite{briesemeister-et-al:j:yloc}, we replicated the YLoc$^+$ runs using our 5-way splits, and obtained results that closely match those originally reported by Briestmeister et al~\cite{briesemeister-et-al:j:yloc}. (The replicated $F_1$-$label$ score is 0.69 with standard deviation of $\pm 0.01$, compared to YLoc$^+$ reported $F_1$-$label$ score of 0.68, and the replicated accuracy is 0.65 with standard deviation of $\pm 0.01$, compared to YLoc$^+$ reported accuracy of 0.64). The total training time for our system is about 11 hours (wall-clock), when running on a standard Dell Poweredge machine with 32 AMD Opteron 6276 processors. Notably, no optimization or heuristics for improving run time were employed, as this is a one-time training. For the experiments described here, we ran 25 training experiments, through 5 times 5-fold cross validation, where the total run time was about 75 hours (wall clock).

We use in our evaluation the {\it adapted} measures of {\it accuracy} and {\it $F_1$ score} proposed by Tsoumakas~\cite{tsoumakas:j:mulilabel-overview} for evaluating multi-label classification. Some of these measures have also been previously used for multi-location evaluation~\cite{he-et-al:j:immml,briesemeister-et-al:j:yloc}. 
To formally define these measures, let $D$ be a dataset containing $m$ proteins. For a given a protein $P$, let $M^{P} \!=\! \{s_i~|~l^{P_i}\!=\!1 \textnormal{, where $1 \leq i \leq q$}\}$ be the set of locations to which protein $P$ localizes, and let $\hat{M}^{P} \!=\! \{s_i~|~\hat{l}^{P_i}\!=\!1 \textnormal{, where $1 \leq i \leq q$}\}$ be the set of locations that a classifier predicts for protein $P$, where $\hat{l}^{P_i}$ is the 0/1 prediction obtained (as described in Section~\ref{sec:methods}). The multi-label accuracy and the multi-label $F_1$ score are defined as:  
\begin{align*}
Acc \!\!=\!\! \frac{1}{m}\! \sum_{j=1}^m \!\!\frac{|M^j \cap \hat{M^j}|}{|M^j \cup \hat{M^j}|} \text{ and }
F_1 \!\!=\!\! \frac{1}{m}\! \sum_{j=1}^m\!\! \frac{2|M^j \cap \hat{M^j}|}{|M^j| + |\hat{M^j}|}.
\end{align*}

Adapted measures of Precision and Recall, denoted $Pre_{s_i}$ and $Rec_{s_i}$ are used to evaluate how well our system classifies proteins as localized or not localized to any single location $s_i$~\cite{briesemeister-et-al:j:yloc}. The \emph{Multilabel-Precision} is: 
\begin{align*}
Pre_{s_i} \!=\! \frac{1}{|\{P \in D~|~s_i \in \hat{M^P}\}|}\! \sum_{P \in D~|~s_i \in \hat{M^P}}\! \frac{|M^P \cap \hat{M^P}|}{|\hat{M^P}|}, 
\end{align*}
and the \emph{Multilabel-Recall} is: 
\begin{align*}
Rec_{s_i} \!=\! \frac{1}{|\{P \in D~|~s_i \in M^P\}|}\! \sum_{P \in D~|~s_i \in M^P}\! \frac{|M^P \cap \hat{M^P}|}{|M^P|}.
\end{align*}
Note that $Pre_{s_i}$ captures the ratio of the number of correctly predicted multiple locations to the total number of multiple locations predicted, and $Rec_{s_i}$ captures the ratio of the number of correctly predicted multiple locations to the number of original multiple locations, for all the proteins that co-localize to location $s_i$. Therefore, high values of these measures for proteins that co-localize to the location $s_i$ indicate that the sets of predicted locations that include location $s_i$ are predicted correctly.
Additionally, the $F_1$-{\it label} score used by Briesemeister et al.~\cite{briesemeister-et-al:j:yloc} to evaluate the performance of multi-location predictors is computed as follows: 
\begin{align*}
F_1\text{-{\it label}}\!=\! \frac{1}{|S|}\! \sum_{s_i \in S}\! \frac{2 \times Pre_{s_i} \times Rec_{s_i}}{Pre_{s_i} + Rec_{s_i}}.
\end{align*}

Finally, to evaluate the correctness of predictions made for each location $s_i$, we use the \emph{standard precision} and \emph{recall} measures, denoted by $Pre$-$Std_{s_i}$ and $Rec$-$Std_{s_i}$ (e.g. \cite{shatkay-et-al:j:sherloc}) and defined as: 
\begin{align*}
Pre\textnormal{-}Std_{s_i} = \frac{TP}{TP+FP} \textnormal{ and } Rec\textnormal{-}Std_{s_i} = \frac{TP}{TP+FN},
\end{align*} 
where $TP$ (\emph{true positives}) denotes the number of proteins that localize to $s_i$ and are predicted to localize to $s_i$, $FP$ (\emph{false positives}) denotes the number of proteins that do not localize to $s_i$ but are predicted to localize to $s_i$, and $FN$ (\emph{false negatives}) denotes the number of proteins that localize to $s_i$ but are not predicted to localize to $s_i$. 

\begin{table}[!htb]
%\vspace{-0.5cm}
\centering
\caption{Multi-location prediction results, averaged over 25 runs of 5-fold cross-validation, for multi-localized proteins only, using our system, YLoc$^{+}$\cite{briesemeister-et-al:j:yloc}, Euk-mPLoc~\cite{chou-et-al:j:eukmploc}, WoLF PSORT~\cite{horton-et-al:j:wolfpsort}, and KnowPred~\cite{lin-et-al:j:knowpred}. The $F_1$-{\it label} score and $Acc$ measures shown for all the systems except for ours are taken directly from \emph{Table 3} in the paper by Briesemeister et al.~\cite{briesemeister-et-al:j:yloc}. Standard deviations are provided for our system (not available for other systems). \vspace{-1ex}
\label{tab:multi-perf}}
\resizebox{1.0\columnwidth}{!}{
\begin{tabular}{!{\vrule width 1pt}c!{\vrule width 1pt}c!{\vrule width 1pt}c!{\vrule width 1pt}c!{\vrule width 1pt}c!{\vrule width 1pt}c!{\vrule width 1pt}} \hhline{|-|-|-|-|-|-|} \hline
 & {\cellcolor{gray!25}}{\bf Our system} & {\bf $\textbf{YLoc}^+$\cite{briesemeister-et-al:j:yloc}} & {\bf Euk-mPLoc~\cite{chou-et-al:j:eukmploc}} & {\bf WoLF PSORT~\cite{horton-et-al:j:wolfpsort}} & {\bf KnowPred~\cite{lin-et-al:j:knowpred}} \\ \hhline{|-|-|-|-|-|-|} \hline
{\bf $\boldsymbol{F_1}$-$\boldsymbol{label}$} & {\cellcolor{gray!25}}0.66 ($\pm$ 0.02) & 0.68 & 0.44 & 0.53 & 0.66 \\ \hhline{|-|-|-|-|-|-|} \hline
{\bf $\boldsymbol{Acc}$} & {\cellcolor{gray!25}}0.63 ($\pm$ 0.01) & 0.64 & 0.41& 0.43 & 0.63 \\ \hhline{|-|-|-|-|-|-|} \hline
\end{tabular}
} 
%\vspace{-0.5cm}
\end{table}
\begin{table}[!htb]
%\vspace{-0.2cm}
\centering
\caption{Multi-location prediction results, averaged over 25 runs of 5-fold cross-validation, for the combined set of single- and multi-localized proteins, using our system. The table shows the $F_1$ score, the $F_1$-{\it label} score, and the accuracy ($Acc$) obtained for SVMs without using location inter-dependencies and for our system which uses location inter-dependencies. Standard deviations are shown in parentheses. \vspace{-1ex}
\label{tab:perf}}
\resizebox{0.85\columnwidth}{!}{
\begin{tabular}{!{\vrule width 1pt}c!{\vrule width 1pt}c!{\vrule width 1pt}c!{\vrule width 1pt}c!{\vrule width 1pt}} \hhline{|-|-|-|-|} \hline
 & $\boldsymbol{F_1}$ & $\boldsymbol{F_1}$-$\boldsymbol{label}$ & $\boldsymbol{Acc}$  \\ \hhline{|-|-|-|-|} \hline
{\bf SVMs (without using dependencies)} & 0.77 ($\pm$ 0.01) & 0.67 ($\pm$ 0.02) & 0.72 ($\pm$ 0.01)  \\ \hhline{|-|-|-|-|} \hline
{\bf \cellcolor{gray!25}Our system (using dependencies)} & \cellcolor{gray!25}0.81 ($\pm$ 0.01) & \cellcolor{gray!25}0.76 ($\pm$ 0.02) & \cellcolor{gray!25}0.76 ($\pm$ 0.01) \\ \hhline{|-|-|-|-|} \hline
\end{tabular}
} 
%\vspace{-0.5cm}
\end{table}
\begin{table}[!htb]
%\vspace{-0.2cm}
\centering
\caption{Multi-location prediction results, per location, averaged over 25 runs of 5-fold cross-validation, for the combined set of single- and multi-localized proteins. Results are shown for the five locations $s_i$ that have the largest number of associated proteins (the number of proteins per location is given in parenthesis): cytoplasm (cyt), extracellular space (ex), nucleus (nuc), membrane (mem), and mitochondrion (mi). The table shows the measures (\emph{standard precision} ($Pre$-$Std_{s_i}$) and \emph{recall} ($Rec$-$Std_{s_i}$), and \emph{Multilabel-Precision} ($Pre_{s_i}$) and \emph{Multilabel-Recall} ($Rec_{s_i}$)), obtained for SVMs without using location inter-dependencies and for our system by using location inter-dependencies. The highest values  between the two methods are shown in boldface. Standard deviations are shown in parentheses. \vspace{-1ex}
\label{tab:label-perf}}
\resizebox{1.0\columnwidth}{!}{
\begin{tabular}{!{\vrule width 1pt}c!{\vrule width 1pt}c!{\vrule width 1pt}c!{\vrule width 1pt}c!{\vrule width 1pt}c!{\vrule width 1pt}c!{\vrule width 1pt}} \hhline{|-|-|-|-|-|-|}\hhline{|-|-|-|-|-|-|} \hline
 & {\bf cyt (3785)} & {\bf ex (1405)} & {\bf nuc (2952)} & {\bf mem (1824)} & {\bf mi (870)} \\ \hhline{|-|-|-|-|-|-|}\hhline{|-|-|-|-|-|-|} \hline
$\boldsymbol{Pre}$-$\boldsymbol{Std_{s_i}}$ {\bf (SVMs)} & 0.84 ($\pm$ 0.01) & 0.87 ($\pm$ 0.02) & 0.79 ($\pm$ 0.02) & {\bf 0.93 ($\pm$ 0.01)} & {\bf 0.90 ($\pm$ 0.03)} \\ \hhline{|-|-|-|-|-|-|}\hhline{|-|-|-|-|-|-|} \hline
\cellcolor{gray!25}$\boldsymbol{Pre}$-$\boldsymbol{Std_{s_i}}$ {\bf (Our system)} & \cellcolor{gray!25}{\bf 0.84 ($\pm$ 0.01)} & \cellcolor{gray!25}{\bf 0.91 ($\pm$ 0.02)} & \cellcolor{gray!25}{\bf 0.79 ($\pm$ 0.03)} & \cellcolor{gray!25}0.90 ($\pm$ 0.01) & \cellcolor{gray!25}0.87 ($\pm$ 0.03) \\ \hhline{|-|-|-|-|-|-|}\hhline{|-|-|-|-|-|-|} \hline
$\boldsymbol{Rec}$-$\boldsymbol{Std_{s_i}}$ {\bf (SVMs)} & 0.85 ($\pm$ 0.01) & 0.64 ($\pm$ 0.02) & 0.72 ($\pm$ 0.02) & 0.79 ($\pm$ 0.02) & 0.62 ($\pm$ 0.03)  \\ \hhline{|-|-|-|-|-|-|} \hhline{|-|-|-|-|-|-|} \hline
\cellcolor{gray!25}$\boldsymbol{Rec}$-$\boldsymbol{Std_{s_i}}$ {\bf(Our system)} & \cellcolor{gray!25}{\bf 0.86 ($\pm$ 0.01)} & \cellcolor{gray!25}{\bf 0.65 ($\pm$ 0.02)} & \cellcolor{gray!25}{\bf 0.74 ($\pm$ 0.03)} & \cellcolor{gray!25}{\bf 0.80 ($\pm$ 0.02)} & \cellcolor{gray!25}{\bf 0.66 ($\pm$ 0.03)}  \\ \hhline{|-|-|-|-|-|-|}\hhline{|-|-|-|-|-|-|} \hline
$\boldsymbol{Pre_{s_i}}$ {\bf (SVMs)} & {\bf 0.82 ($\pm$ 0.01)} & 0.89 ($\pm$ 0.02) & 0.83 ($\pm$ 0.01) & {\bf 0.92 ($\pm$ 0.01)} & 0.87 ($\pm$ 0.03)   \\ \hhline{|-|-|-|-|-|-|}\hhline{|-|-|-|-|-|-|} \hline
\cellcolor{gray!25}$\boldsymbol{Pre_{s_i}}$ {\bf(Our system)} & \cellcolor{gray!25}0.81 ($\pm$ 0.02) & \cellcolor{gray!25}{\bf 0.91 ($\pm$ 0.02)} & \cellcolor{gray!25}{\bf 0.83 ($\pm$ 0.01)} & \cellcolor{gray!25}0.90 ($\pm$ 0.01) & \cellcolor{gray!25}{\bf 0.89 ($\pm$ 0.02)}   \\ \hhline{|-|-|-|-|-|-|}\hhline{|-|-|-|-|-|-|} \hline
$\boldsymbol{Rec_{s_i}}$ {\bf (SVMs)} & 0.78 ($\pm$ 0.01) & 0.72 ($\pm$ 0.02) & 0.77 ($\pm$ 0.01) & 0.76 ($\pm$ 0.01) & 0.68 ($\pm$ 0.02) \\ \hhline{|-|-|-|-|-|-|}\hhline{|-|-|-|-|-|-|} \hline
\cellcolor{gray!25}$\boldsymbol{Rec_{s_i}}$ {\bf(Our system)} & \cellcolor{gray!25}{\bf 0.80 ($\pm$ 0.01)} & \cellcolor{gray!25}{\bf 0.74 ($\pm$ 0.02)} & \cellcolor{gray!25}{\bf 0.78 ($\pm$ 0.02)} & \cellcolor{gray!25}{\bf 0.78 ($\pm$ 0.01)} & \cellcolor{gray!25}{\bf 0.73 ($\pm$ 0.02)} \\ \hhline{|-|-|-|-|-|-|}\hhline{|-|-|-|-|-|-|} \hline
\end{tabular}
} 
\vspace{-0.2cm}
\end{table}

\subsection{Classification Results}
\label{subsec:pred-res}
Table~\ref{tab:multi-perf} shows the $F_1$-{\it label} score and the accuracy for our system in comparison to those obtained by other predictors (as reported by Briesemeister et al.~\cite{briesemeister-et-al:j:yloc}, \emph{Table 3} there, using the same set of multi-localized proteins and evaluation measures. While the table shows that our system has a slightly lower performance than YLoc$^{+}$, the differences in the values are not statistically significant, and the overall performance level is comparable. Thus our approach performs as effectively as current top-systems, while having the advantage of directly capturing inter-dependencies among locations in a generalizable manner (that is, without introducing a new location-class for each new location-combination).

Table~\ref{tab:perf} shows the $F_1$ score, the $F_1$-{\it label} score, and the accuracy obtained by the individual SVM classifiers (used for computing estimates of location indicators) without using location inter-dependencies compared with the corresponding values obtained by our system by using location inter-dependencies, on the combined dataset of both single- and multi-localized proteins. All the scores obtained by using inter-dependencies are significantly higher than those obtained by using SVMs alone without utilizing inter-dependencies. These differences are highly statistically significant ($p \ll 0.001$), as measured using the 2-sample t-test~\cite{degroot:b:prob-stat}.

Table~\ref{tab:label-perf} shows the prediction results obtained by our system for the five locations that have the largest number of associated proteins: cytoplasm (cyt), extracellular space (ex), nucleus (nu), membrane (mem), and mi (mitochondrion), on the combined dataset of both single- and multi-localized proteins. For each location $s_i$, we show the \emph{standard precision} ($Pre$-$Std_{s_i}$) and \emph{recall} ($Rec$-$Std_{s_i}$) as well as the \emph{Multilabel-Precision} ($Pre_{s_i}$) and \emph{Multilabel-Recall} ($Rec_{s_i}$). The table shows values for each of the measures obtained by SVMs without using location inter-dependencies and by our system using location inter-dependencies. When using inter-dependencies, we note that for all locations the \emph{Multilabel-Recall} ($Rec_{s_i}$) increases (in some cases statistically significantly); while for a few locations (such as cytoplasm and membrane) the \emph{Multilabel-Precision} ($Pre_{s_i}$) decreases, the decrease is not statistically significant. For instance, when classifying using SVMs without using inter-dependencies $Rec_{cyt}$ is 0.78 and $Rec_{mem}$ is 0.76, while when incorporating the inter-dependencies the recall is 0.80 and 0.78, respectively. Even for locations with fewer associated proteins, e.g. peroxisome, (157 proteins), the \emph{Multilabel-Recall} increases from 0.37 using simple SVMs to 0.65 using our classifier. This demonstrates the advantage of using location inter-dependencies for predicting protein locations, not just for locations that have a large number of associated proteins but also for locations that have relatively few associated proteins. 
\section{Discussion and Future Work}
\label{sec:discussion}
We presented a new way to use a collection of Bayesian network classifiers taking advantage of location inter-dependencies to provide a generalizable method for predicting possible multiple locations of proteins. The results demonstrate that the performance of our preliminary system is comparable to the best current multi-location predictor YLoc$^{+}$\cite{briesemeister-et-al:j:yloc}, which indirectly addresses dependencies by creating a class for each multi-location combination. Our results also show that utilizing inter-dependencies significantly improves the performance of the location prediction system, with respect to SVM classifiers that do not use any inter-dependencies. 

In most biological applications that have used Bayesian networks so far (e.g. \cite{friedman-et-al:j:bn-exprdata,segal-et-al:j:prob-geneexp,lee-et-al:j:bntagger}), the variable-space typically corresponds to genes or SNPs which is a very large space and necessitates the use of strong simplifying assumptions and many heuristics. In contrast, we note that predicting multiple locations for proteins involves a significantly smaller number of variables (as the number of subcellular components and the number of features for representing proteins are relatively small), making this task ideally suitable for the use of Bayesian networks. 

The study presented here is a first investigation into the benefit of directly modeling and using location inter-dependencies. In order to obtain initial estimates for location values, we used a simple SVM classifier, and location inter-dependencies were only learned based on these values. While the results have already shown much improvement with respect to the baseline SVM classifiers, we believe that a better approach would be to simultaneously learn a Bayesian network while estimating the location values using methods such as expectation maximization. 

We note that although the dataset we use contains the most extensive available collection of multi-localized proteins, several subcellular locations are not represented in the dataset at all due to the low number of proteins associated with them. Similarly, there is not enough data pertaining to proteins that are localized to more than two locations. We are in the process of constructing a set of multi-localized proteins that will be used in future work to test the performance of our system on novel, and more complex, combinations. We also plan to develop improved approaches for learning models of location inter-dependencies from the available data.

\vspace{0.2cm}
\noindent
{\bf Acknowledgments:} We are grateful to S. Briesemeister for so readily providing us with information about the implementation and testing of YLoc$^{+}$.
\bibliographystyle{main}
\bibliography{main}
\end{document}